\documentclass[letterpaper, 10 pt, journal, twoside]{IEEEtran}

\ifCLASSINFOpdf
\else
\fi

\usepackage{color,array}

\usepackage{graphicx}
\usepackage{amssymb}
\usepackage{amsmath}
\usepackage{caption}
\usepackage{subcaption}
\usepackage{cite}
\usepackage{amsmath,amssymb,amsfonts}
\usepackage{hyperref}

\begin{document}
%
\title{Radar-based Materials Classification Using Deep Wavelet Scattering Transform: A Comparison of Centimeter vs. Millimeter Wave Units}
%
%
%

\author{Rami N. Khushaba, \IEEEmembership{Senior Member, IEEE}, Andrew J. Hill, \IEEEmembership{Member, IEEE}
\thanks{Manuscript received: September, 9, 2021; Revised: December, 7, 2021; Accepted January, 10, 2022.}
\thanks{This paper was recommended for publication by Prof. Markus Vincze upon evaluation of the Associate Editor and Reviewers' comments.}
\thanks{The authors have no affiliation with Vayyar Imaging, or its partners involved in design, manufacturing or sale of the radars tested in this paper.}
\thanks{All authors are within the Australian Centre for Field Robotics, The University of Sydney. 8 Little Queen street, Chippendale, NSW 2008. (e-mail: Rami.Khushaba@Sydney.edu.au; Andrew.Hill@Sydney.edu.au).}
\thanks{Digital Object Identifier (DOI): see top of this page.}
}
%
%

\markboth{IEEE Robotics and Automation Letters. Preprint Version. Accepted January, 2022}
{Khushaba \MakeLowercase{\textit{et al.}}: Radar-based Materials Classification Using Deep Wavelet Scattering Transform} 
%



\maketitle

\begin{abstract}
Radar-based materials detection received significant attention in recent years for its potential inclusion in consumer and industrial applications like object recognition for grasping and manufacturing quality assurance and control. Several radar publications were developed for material classification under controlled settings with specific materials' properties and shapes. Recent literature has challenged the earlier findings on radars-based materials classification claiming that earlier solutions are not easily scaled to industrial applications due to issues such as the analog-to-digital converters' high sensitivity to target aspect angle, noise fluctuations due to temperature and other external conditions and sensor orientation. Published experiments on the impact of the aforementioned factors on the robustness of the extracted radar-based traditional features have already demonstrated that the application of deep neural networks can mitigate, to some extent, the impact to produce a viable solution. However, previous studies lacked an investigation of the usefulness of lower frequency radar units, specifically $<10$GHz, against the higher range units around and above 60GHz. To address the aforementioned investigation, this research considers two radar units with different frequency ranges: the Walabot-3D (6.3-8 GHz) cm-wave and IMAGEVK-74 (62-69 GHz) mm-wave imaging units by Vayyar Imaging. A comparison is presented on the applicability of each unit for material classification. This work also extends upon previous efforts, by applying deep wavelet scattering transform for the identification of different materials based on the reflected signals received by these units. In the wavelet scattering feature extractor, data is propagated through a series of wavelet transforms, nonlinearities, and averaging to produce low-variance representations of the reflected radar signals. This work is unique in terms of the comparison of the utilized radar units and algorithms in material classification and includes real-time demonstrations that show strong performance by both units, with increased robustness offered by the cm-wave radar unit.

\end{abstract}


%
\IEEEpeerreviewmaketitle

\section{INTRODUCTION}

\IEEEPARstart{W}{ireless} broadband sensing has been utilized as an effective tool for the unique characterization of materials based on their dielectric properties \cite{Ref000,Ref001}. Compared to other properties, permittivity (or dielectric constant) is promising for the implementation of a sensing unit and offers a potential for various applications in medical, biological, and agricultural fields \cite{Ref002,Ref003,Ref004,Ref006}. Ultra-Wideband (UWB) radar signals are characterized for having both high frequency carrier and high bandwidth \cite{Ref007}. This makes the scattered field from the targets when irradiated with UWB pulses highly dependent on the composition and shape of the target, the absorption and scattering properties of the material at the wavelengths used, the refractive index, and thus the specular reflection from the material. However, parameters such as object thickness, dimensions, impurities, and reflections from the surrounding test setup, could also affect the received signals in such sensing and radar applications, which would increase the uncertainty in the detection results. In such a case, the use of supervised machine learning (ML) approaches with numerous samples for the materials under test was recommended for application in radar based materials detection \cite{Ref008}.

For the materials identification problem, a number of research attempts have considered the use of centimeter-wave (3 to 30 GHz) and millimeter-wave (30 to 300 GHz) radar units individually to distinguish a number of materials from the reflected radar signals. For example, the RadarCat system \cite{RdarCat01,RdarCat02} employed Google Soli radar unit (57--64 GHz) to recognize different body parts, everyday objects and materials including transparent materials. While being very successful, the RadarCat system was criticized in that it can not be readily scaled to industrial or consumer applications as the ADC data is very sensitive to target aspect angle, sensor orientation, noise variations due to temperature and other environmental factors \cite{Avik01}. Jonas and Avik \cite{Avik00} extended the RadarCat work by addressing the problem of generalization and demonstrated an approach based on a Siamese Convolutional Neural Network (CNN) for classifying five variations of four materials. This was done while employing Infineon's short-range 60 GHz compact radar sensor (covering 57--64 GHz), achieving an overall accuracy of $99.23\%$. The authors further extended their work by considering the invariance to sensor noise and sensor nonlinearities, achieving a classification accuracy of above $97\%$ while classifying 1875 radar images from ten different materials \cite{Avik01}.

Brook et. al. \cite{Ref009} further proposed the use of a portable laboratory device operating in a frequency-modulated continuous-wave (FMCW) mode over 40 and 90 GHz bandwidths (centered at 100 and 300 GHz respectively), for the detection of defects in manufactured calibration materials. The recent work by Jamali et. al. \cite{Ref010} further looked at using a millimeter-wave radar working in the 75--110 GHz range with different classification models including Support Vector Machines, Multilayer Perceptrons, and Gaussian Process classifiers to characterize dielectric slabs made of different materials and thicknesses. On the other hand, the work by Bouza et. al. \cite{Ref007} demonstrated the feasibility of using a UWB pulse radar with a center frequency of 4.25 GHz to discriminate between groups of 3, 4, and 5 materials using a support vector machine classifier. Unlike previous work, Agresti and Milani \cite{Ref008} employed a portable 3D imaging radar-based system (the Walabot sensor by Vayyar Imaging working in the 6.3-8 GHz frequency range) to acquire three-dimensional radiance maps of various analyzed objects and processed that by using CNNs in order to identify which material the object is made of. However, missing from the literature is direct comparison between many of these units and the justification for using one frequency range/radar unit against another, with the perceived benefits. Additionally, the generalization of the offline findings of many experiments into real-time tests is not yet demonstrated.

This paper aims to extend the work in the literature by 1) considering two radar units with different frequency ranges including the Walabot (3.3--10 GHz US, 6.3--8 GHz EU) and the IMAGEVK-74 (62--69 GHz) to study the potential benefits and performance comparisons between the two units, and by proxy the two frequency bands; 2) employing the Wavelet scattering transform (WST) to generate translation-invariant and deformation-stable representations of the radar signals through cascades of wavelet convolutions with nonlinear modulus and averaging operators to achieve the required robustness; and 3) experimentally demonstrating the real-time performance of this approach with these radar units. Unlike existing solution suggesting deep learning models for materials classification, the approach based on the scattering transform is not iterative (require no training) and can function with a small sample size, while deep learning models require a huge amount of training data and to optimize the different parameters across a significant number of training iterations.

\section{Materials and Experimental Methods}
\begin{figure}[!tb]
\centering 
\includegraphics[width = .41\textwidth]{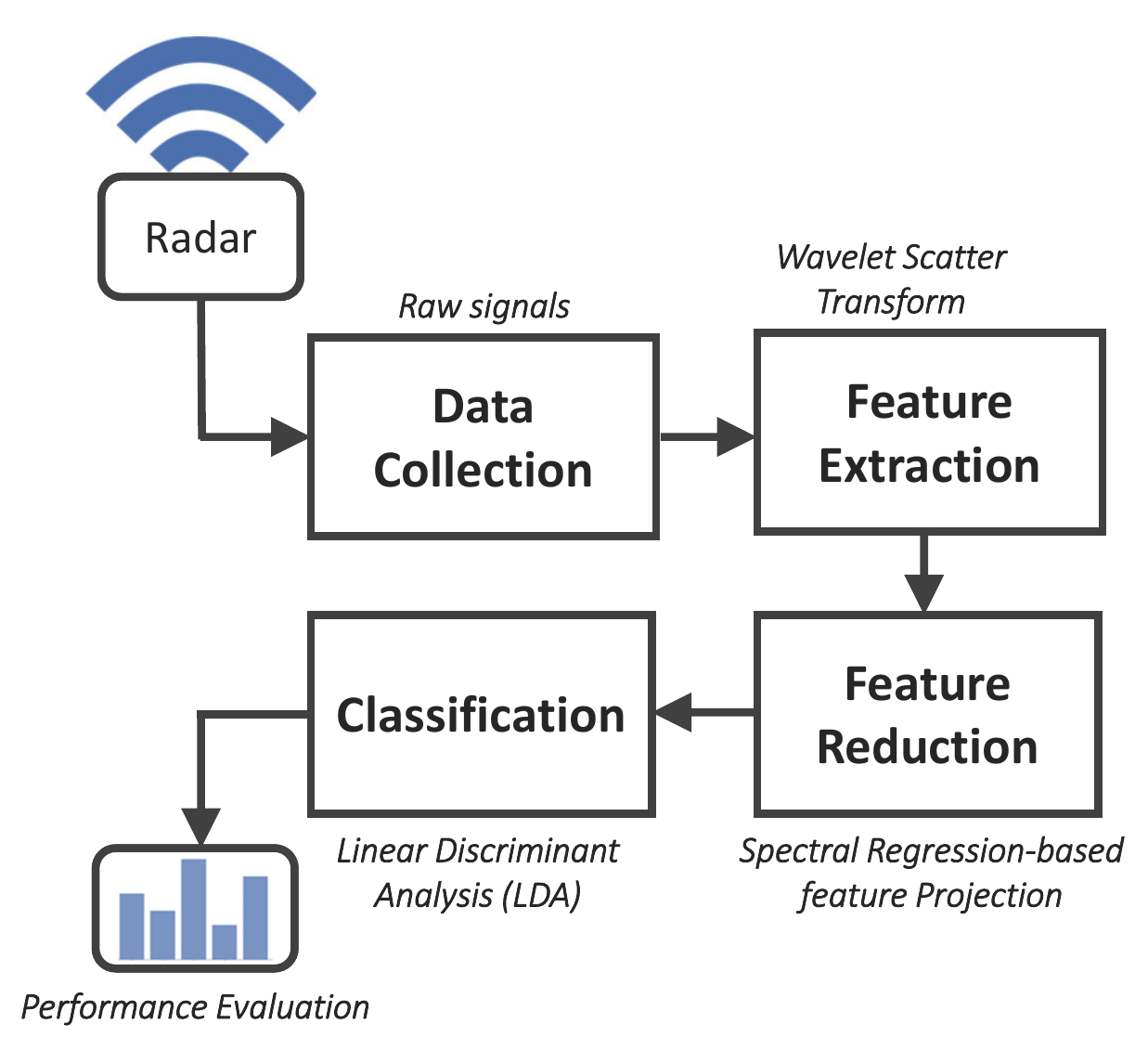}
\caption{Block diagram of the proposed radar-based materials recognition pipeline.}
\label{FigureWST}
\end{figure}

The block diagram of the proposed pipeline is shown in Fig.\ref{FigureWST}. In the following sections, we start first with the description of the radar units employed in this study, followed by the data collection procedure. The paper proceeds then with the description of the wavelet scattering transform, followed by the details of the classification pipeline involving dimensionality reduction and classification methods utilized.

\subsection{Walabot Radar}
\begin{figure}[!tb]
\centering 
\includegraphics[width = 0.26\textwidth]{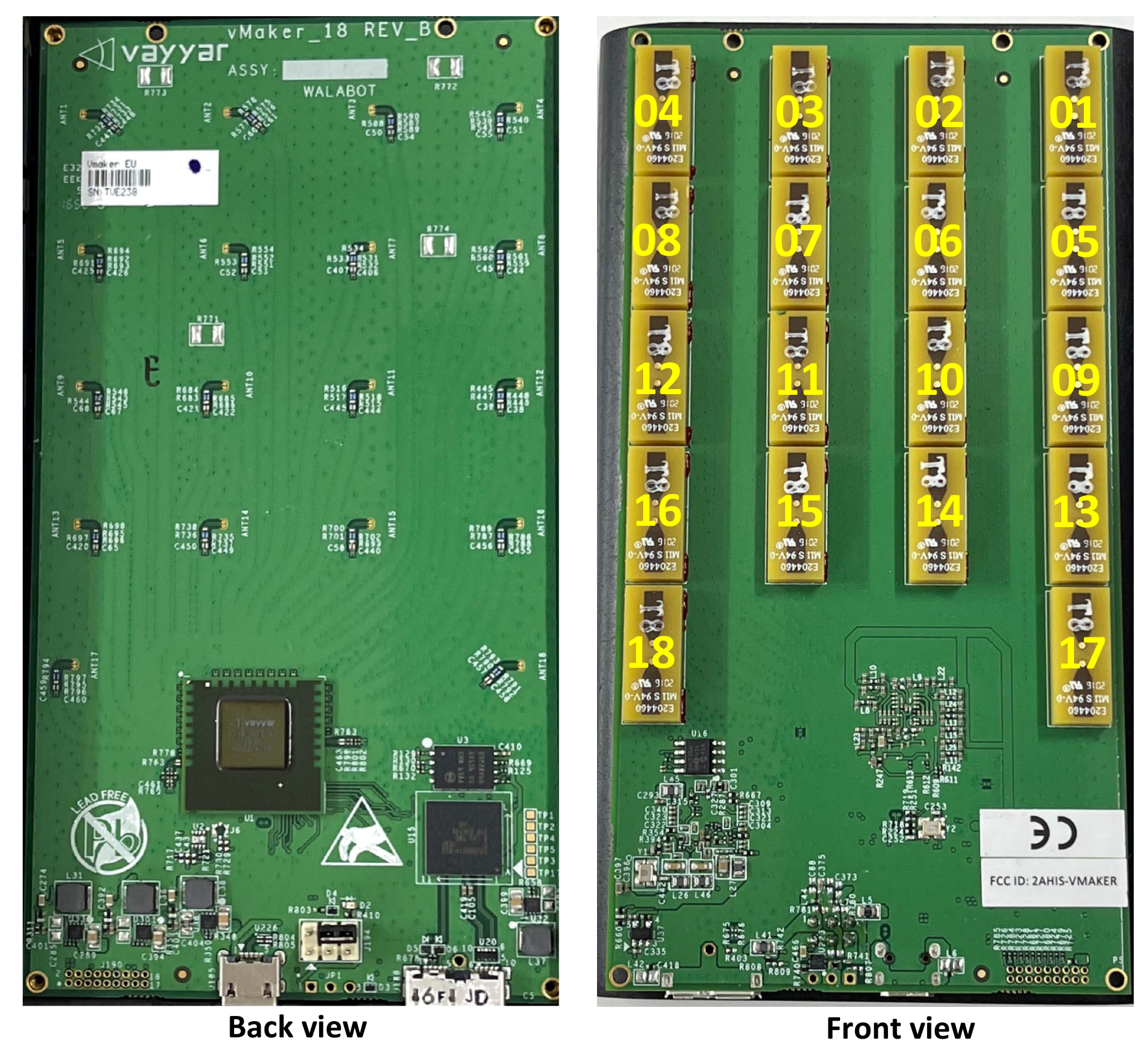}
\caption{Walabot Radar unit from Vayyar Imaging. The left image shows the rear of the board with the VYYR2401 chip, USB controller and micro-USB connectors. The right image shows the antenna array.}
\label{Figure02}
\end{figure}

The Walabot from Vayyar Imaging\footnote{https://vayyar.com/} is a programmable 3D radio frequency (RF) sensor unit operating in the frequency range of 3.3--10 GHz (US model) or 6.3--8 GHz (European model). It employs a 72mm $\times$ 140mm array of 18 antennas (4 transmitting, 14 receiving) to generate a 3D representation from the area facing the antenna array by measuring the strength of the reflected signal (see Fig.\ref{Figure02}). The intensity of each `voxel' in this representation denotes the reflected energy received at a given angle and distance. The average transmit power is below $-41$ dBm/MHz, so it can be safely used in public places without restrictions. A combination of transmitting and receiving antennas are coupled to generate 40 raw signals that are provided by the Walabot SDK. The radiated space is sampled following a spherical coordinate reference system; each sample is rescaled and quantized into an 8-bit integer value\cite{Ref008}.

The implemented API permits collecting three forms of outputs from the Walabot unit: 1) a 2D energy map, with the width and height of this map defined according to user settings in the spherical polar coordinates ($R$; $\theta$; $\phi$); 2) the raw signals associated to the 40 couples of transmitting/receiving antennas, which are sampled at frequency of 100 GHz and represented by arrays of 8192 values in double precision; and 3) a 3D point cloud reporting the reflectance (8 bit integer) of each point localized in spherical polar coordinates ($R$; $\theta$; $\phi$). The Walabot has been utilized in several applications in the literature including the detection of humans and objects through the wall \cite{Walabot001}, heart rate and respiration monitoring \cite{Walabot002,Walabot003}, pose estimation \cite{Walabot005}, activity recognition \cite{Walabot004}, soil moisture monitoring \cite{Walabot006}, and materials identification \cite{Ref008}.

\subsection{IMAGEVK-74 Radar}

\begin{figure}[!tb]
\centering 
\includegraphics[width = 0.45\textwidth]{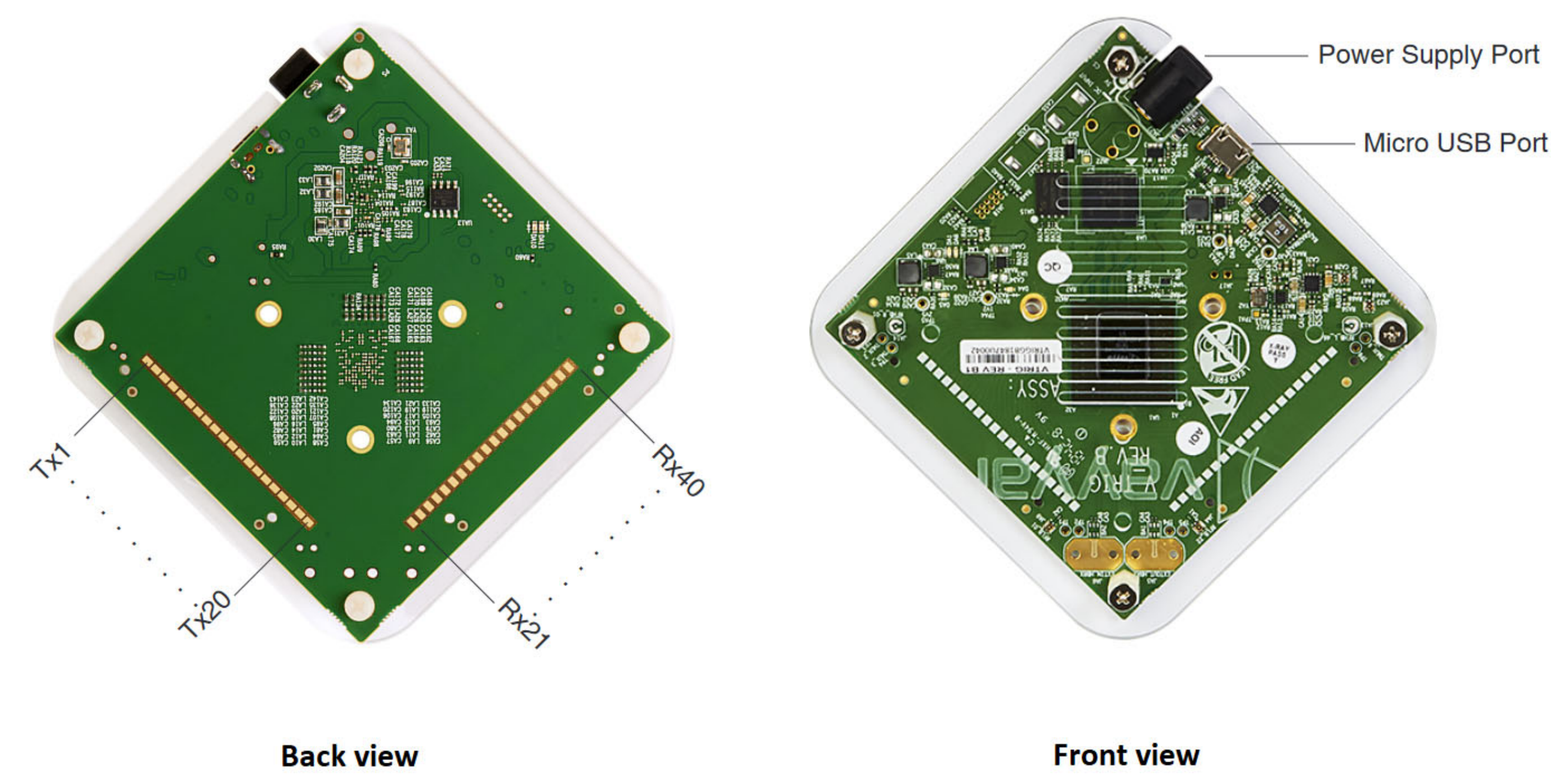}
\caption{IMAGEVK-74 RF sensor unit from Vayyar Imaging. There are 20 transmitting (Tx) and 20 receiving (Rx) antennas, which allows recording of 400 signals from each combination.}
\label{Figure03}
\end{figure}

The IMAGEVK-74 from Vayyar Imaging is a more advanced model than the Walabot, with an inclusion of 20 transmit (Tx) and 20 receive (Rx) on-board antennas that can be configured to transmit and receive signals anywhere within the 62 to 69 GHz range (see Fig.\ref{Figure03}). It provides a great flexibility for hardware developers and researchers with three performance-optimized transmit profiles to adjust receiver resolution and imaging processing time with direct access to the Tx/Rx pair phasors for each swept frequency point. The high-resolution profile uses 20 Tx and 20 Rx antennas making it ideal for high-resolution 3D imaging while generating 400 raw signals (one per combination of Rx and Tx pair). Similarly to Walabot, IMAGEVK-74 transmits the signals with an effective radiated power of less than $-5$ dBm making it safe for use around people. Similarly to the Walabot, the IMAGEVK-74 also offers a variety of outputs as raw signals, and 3D point clouds. To protect the back of the sensor, a 3D printed plastic cover (Acrylonitrile Styrene Acrylate, ASA) was created to shield the sensor from electrostatic charges or other interference while handling and placing the sensor on the different materials as shown in the video supplementary file. Due to the recent release date of IMAGEVK-74, no publications were found using this specific sensor unit making this research the first application of the IMAGEVK-74, to the best of the authors' knowledge.


\subsection{Data Collection}
\begin{figure}[!t]
\centering 
\includegraphics[width = 0.25\textwidth]{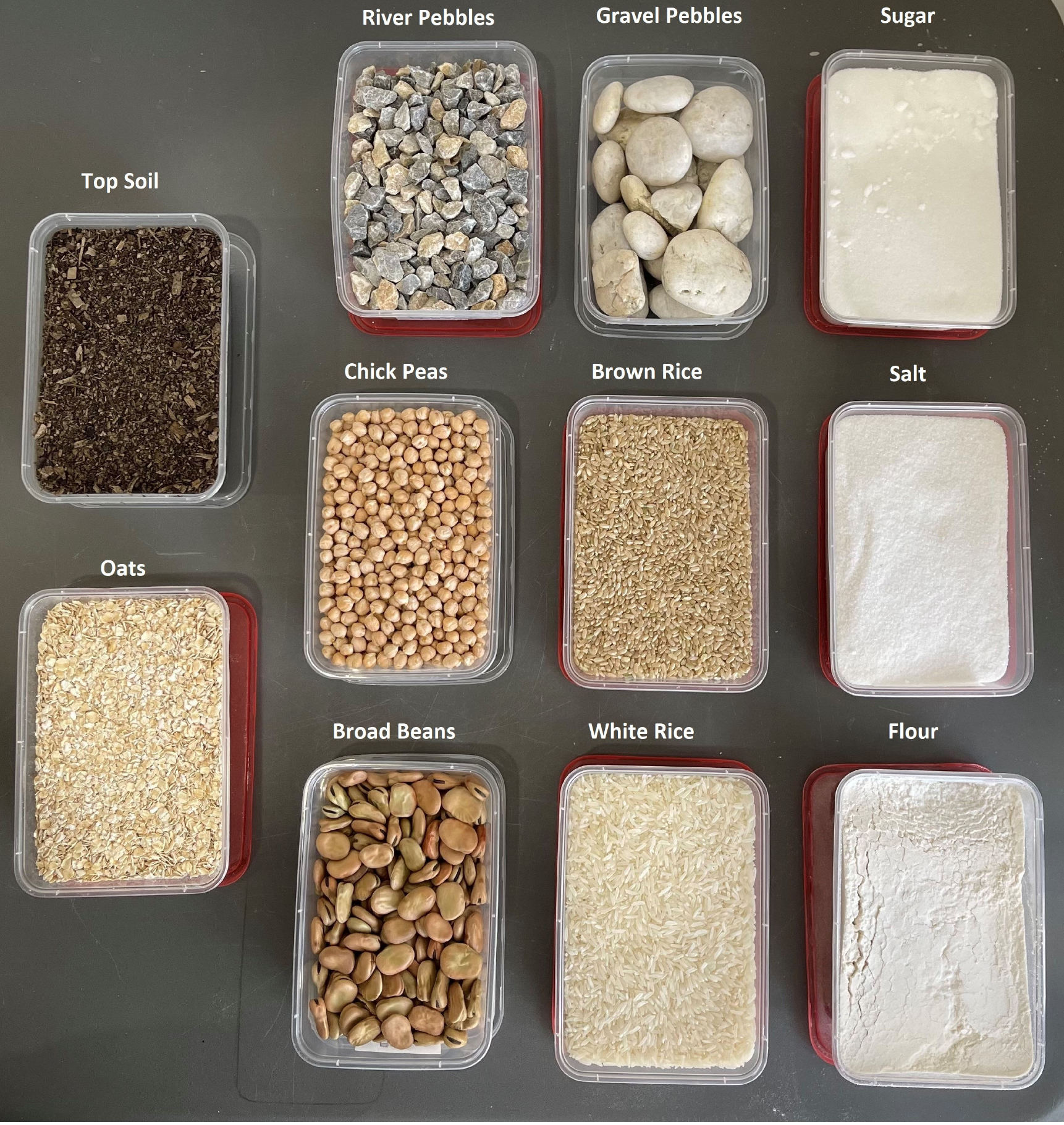}
\caption{Eleven classes of materials considered in this research, in addition to the twelfth class, Air.}
\label{Figure05}
\end{figure}

Eleven unique classes of material were considered in this research, in addition to the case where the radar is not observing at any of the other materials at close proximity (denoted as class "Air"), making a total of twelve classes. When selecting these materials we chose a number of closely related material classes to investigate the true capabilities of the two radar units in differentiating between closely related classes. The chosen classes were 1) River-Pebbles, 2) Gravel-Pebbles, 3) Sugar, 4) Salt, 5) Flour, 6) Broad-Beans, 7) Chickpeas, 8) Top-Soil, 9) Oats, 10) White-Rice, 11) Brown-Rice, and 12) Air. All chosen materials were placed into plastic containers as shown in Fig. \ref{Figure05}, with the two radar units individually placed on top of each of the boxes to identify the corresponding material inside. There were around 300 samples collected by each sensor for each of the materials boxes shown in Fig. \ref{Figure05}. Data collection during the training phase included random sensors' placements on the corresponding boxes (position and orientation), while the testing phase included randomizing the contents of the boxes by shaking (soft materials) and randomly distributing the contents (hard materials). This in turn provides a good amount of variability during testing to make sure that we are not learning the surface profile of the different materials.     


\subsection{Feature Extraction with Wavelet Scattering Transform (WST)}

\begin{figure*}[!tbh]
\centering 
\includegraphics[width = .7\textwidth]{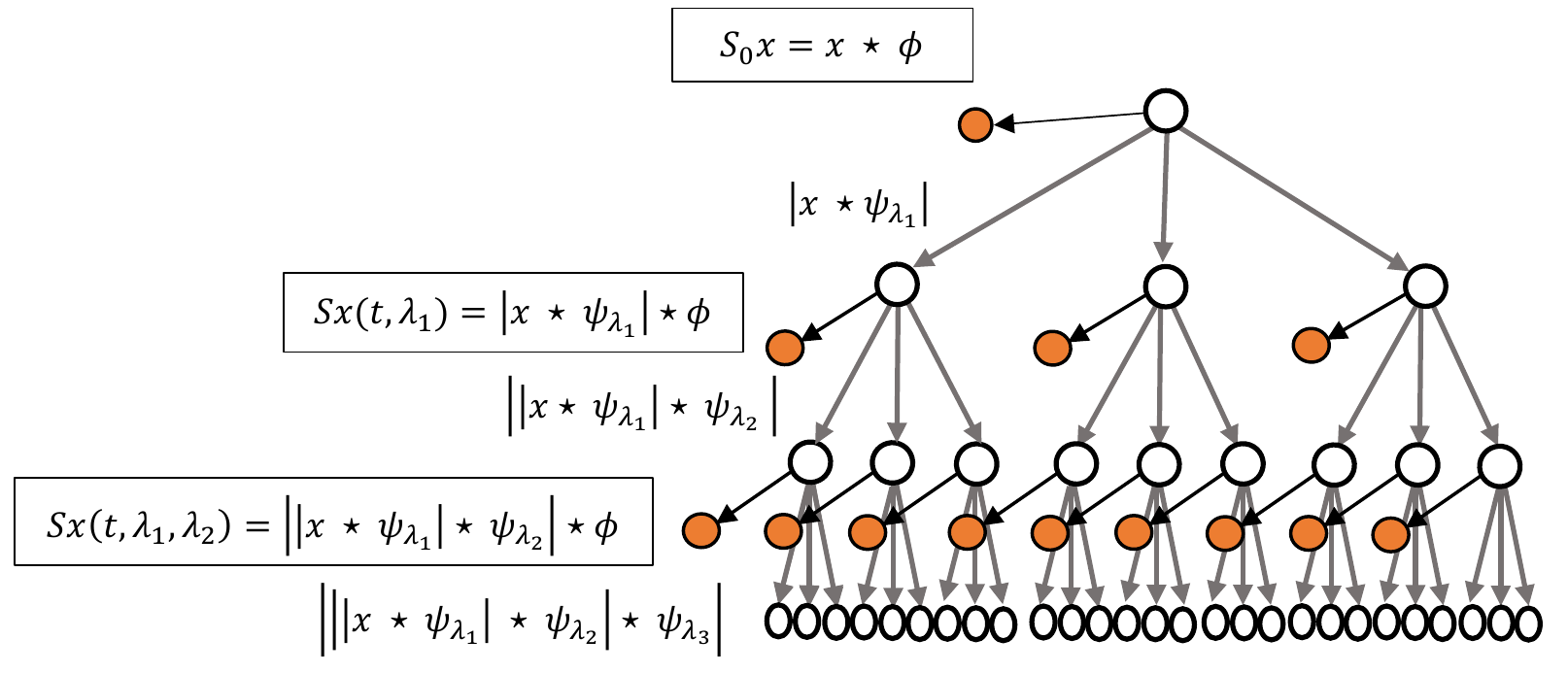}
\caption{A hierarchical representation of the wavelet scattering transform.}
\label{Figure01}
\end{figure*}

A wavelet scattering transform defines a locally translation invariant representation of the underlying signals that is stable to time-warping deformations \cite{Mallat01}. The WST was introduced as a deep representation, obtained by an iterative operator which cascades wavelet convolutions and modulus nonlinearities \cite{Mallat01,Mallat02}. The wavelet transform convolves an input signal $x(t)$ with a filter bank of $\psi_{\lambda_1}$. Such a filter bank is made by dilations of a mother wavelet $\psi(t)$, whose Fourier transform $\widehat{\psi}(w)$ is concentrated over the dimensionless frequency interval [1 - $2^{1/2Q}$; 1 + $2^{1/2Q}$], with $Q$ being the quality factor (the number of wavelet filters per octave for each filter bank). A family of bandpass filters centered at frequencies $\lambda_1 = 2^{j1 + \frac{\chi}{Q}}$ is defined by dilations of the mother wavelet, with the indices of $j1 \in \mathbb{Z}$ and $\chi \in {1...Q}$ respectively denoting the octave and chroma.

\begin{equation}
\widehat{\psi}_{\lambda_1}(w) = \widehat{\psi}(\lambda^{-1}w) \quad \textrm{i.e.,} \quad \psi_{\lambda_1}(w) = \lambda_1 \psi(\lambda_1t).
\end{equation}

A scalogram matrix is then generated by applying the complex modulus to all wavelet convolutions, with a frequential axis that is uniformly sampled by the binary logarithm log$\lambda_1$.

\begin{equation}
x_1(t,\textrm{log}\lambda_1) = \left|x \star \psi_{\lambda_1} \right| \quad \textrm{for all } \lambda_1 > 0,
\end{equation}

\noindent where $\star$ is the convolution operator. The energy of $x(t)$ is localized by the scalogram $x_1$ around frequencies $\lambda_1$ over durations of $2Q\lambda^{-1}_1$ \cite{Mallat03}. The WST coefficients are obtained by averaging the wavelet modulus coefficients with a low-pass filter $\phi(t)$ of size $T$, which ensures local invariance to time-shifts. 

\begin{equation}
S_1x(t,\textrm{log}\lambda1) = x_1 \star \phi_T = \left|x \star \psi_{\lambda_1} \right| \star \phi_T.
\label{equ3}
\end{equation}

At the zero order, a single coefficient is generated by $S_0x(t)$ = $x \star \phi(t)$, as these have very low energy at low frequencies \cite{Mallat01}. As the low-pass filtering process removes all high-frequencies, these are then recovered by a wavelet modulus transform as the time scattering transform also convolves $x_1$ with a second filterbank of wavelets $\psi_{\lambda_2}$ and applies complex modulus to get

\begin{equation}
x_2(t,\textrm{log}\lambda_1,\textrm{log}\lambda_2) = \left| x_1 \star \psi_{\lambda_2} \right| = \left| \left| x \star \psi_{\lambda_1} \right| \star \psi_{\lambda_2} \right|.
\end{equation}

Similarly to Eq.\ref{equ3}, an invariance to translation in time up to $T$ is achieved by averaging. Hence, the second order coefficients, capturing the high-frequency amplitude modulations occurring at each frequency band of the first layer, are obtained by

\begin{equation}
S_2x(t,\textrm{log}\lambda_1,\textrm{log}\lambda_2)= x_2 \star \phi = \left| \left| x \star \psi_{\lambda_1} \right| \star \psi_{\lambda_2} \right| \star \phi_T. 
\end{equation}

\noindent with the wavelet $\psi_{\lambda_2}$ having an octave resolution of $Q_2$, which maybe different from $Q_1$. To define wavelets that are suitable to characterize transients and attacks, a more narrow time support is usually defined by setting $Q_2$ = 1. In this manner, a sparse representation is acquired which leads to concentrating the signal information over as few wavelet coefficients as possible. These coefficients are averaged by the low pass filter $\phi$, which ensures local invariance to time-shifts, as with the first-order coefficients \cite{WST01}. Iterating the aforementioned process defines scattering coefficients at any order $m$ (see Fig.\ref{Figure01} for an illustration of the WST representation). For any $m > 1$, iterated wavelet modulus convolutions are written as:

\begin{equation}
U_mx(t,\textrm{log}\lambda_1,...,\textrm{log}\lambda_m) = \left| \left|\left| x \star \psi_{\lambda_1} \right| \star ...\right| \star \psi_{\lambda_m(t)} \right|,
\end{equation}

\noindent with the $m$th-order wavelets $\psi_{\lambda_m}$ have an octave resolution $Q_m$. The application of $\phi$ on $U_mx$ gives scattering coefficinets of order $m$:

\begin{equation}
\begin{split}
S_mx(t,\textrm{log}\lambda_1,...,\lambda_m) & = \left| \left|\left| x \star \psi_{\lambda_1} \right| \star ... \right| \star \psi\lambda_m\right| \star \phi(t) \\
& = U_m x(.,\textrm{log}\lambda_1,...,\lambda_m) \star \phi(t).
\end{split}
\end{equation}

For a scattering decomposition of maximal order $l$, the final scattering vector aggregates all scattering coefficients for $0 \le m \le l$. 


The WST enables the derivation of low-variance features from real-valued time-series and image data. It provides an invariant and stable signal representation which made it suitable for a wide range of signal processing and machine/deep learning applications that the literature investigated already, including (but not limited to): musical genre and phone segment classification \cite{Mallat01}, handwritten digit recognition \cite{Mallat04, Mallat05}, indoor fingerprint localization \cite{WST02}, texture discrimination \cite{WST041}, and audio classification \cite{WST04}. High-performance Python implementations of the scattering transform in 1D, 2D, and 3D that are compatible with modern deep learning frameworks, like PyTorch and TensorFlow/Keras were also recently made available on both CPUs and GPUs, the latter offering a significant speedup over the former, with advanced implementations supporting highly optimized applications \cite{Kymatio}.

Once the WST is applied on the raw signals collected from the radar units, a number of features would be generated to be submitted to the classification system to produce a decision about the classes that the materials under focus belong to. The cardinality of the WST feature set would depend on the number of raw radar signals collected by the two radar units considered in this research and processed by WST, which is a factor that depends on the selected radar operation profile mentioned in the earlier sections.

\subsection{Feature Reduction and Pattern Classification}
The number of time-domain signals and output formats extracted from each of the radar units is dependent on the selected operating profile. The Walabot offers short-range imaging, sensor, sensor narrow and tracker; the IMAGEVK-74 offers low, medium and high resolution profiles. Given the number of signals generated by these units under the different profiles, the application of WST for feature extraction from these signals can end up with a large number of generated features. The dimensionality of the generated feature set was then reduced using the Spectral Regression (SR) feature projection method \cite{SRDA}, to $c-1$ features, with $c$ being the number of classes in the corresponding dataset. For classification, the classical Linear Discriminant Analysis (LDA) classifier was utilized mainly due to its simplicity and suitability for real-time implementations thanks to its low computational cost.

\section{Experimental Results}

\subsection{Offline Classification Results}
\begin{figure*}[h!]
    \centering
    \begin{subfigure}[b]{0.5\textwidth}
        \centering
        \includegraphics[width=7.2cm]{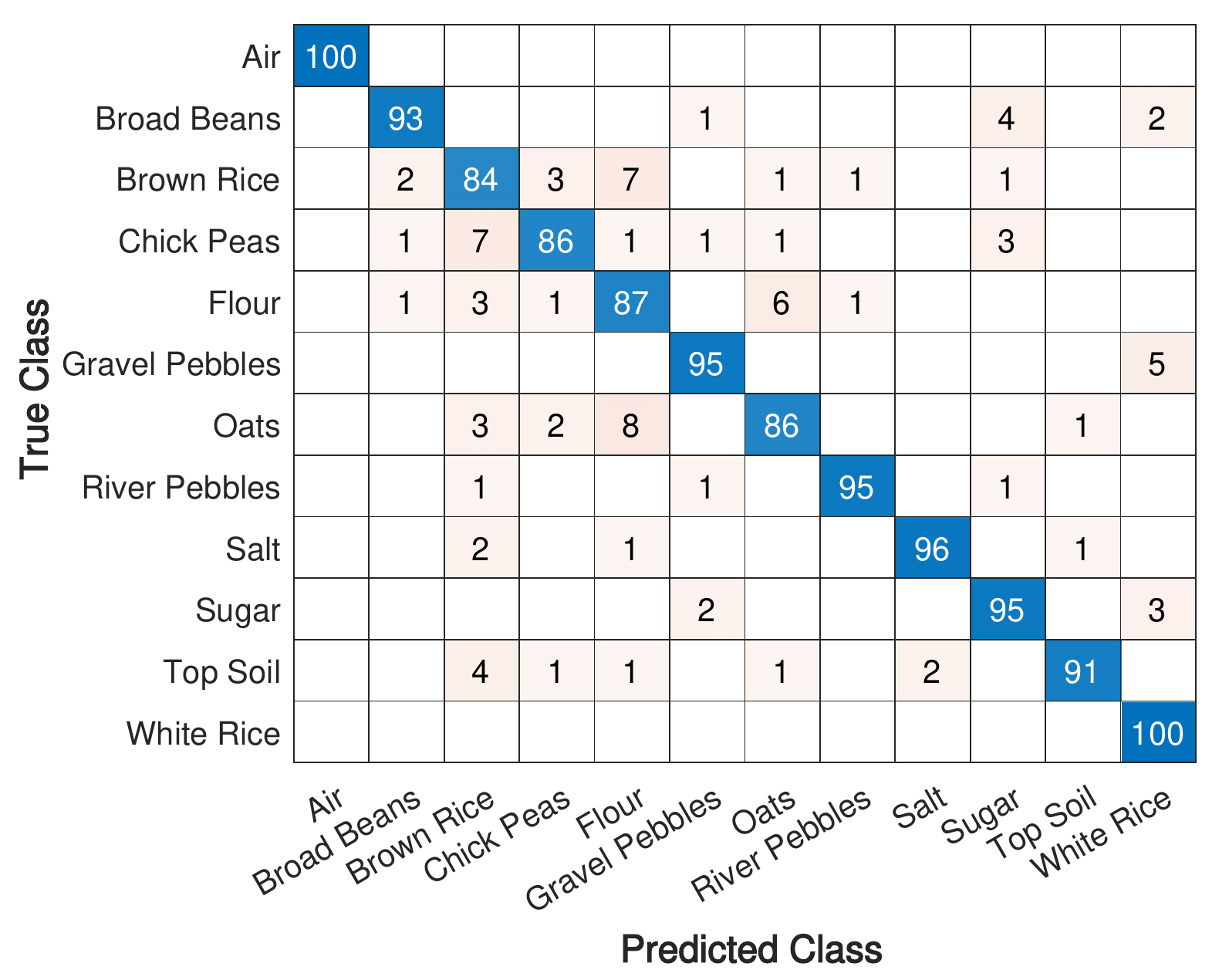}
        \caption{Walabot}
    \end{subfigure}%
    ~ 
    \begin{subfigure}[b]{0.5\textwidth}
        \centering
        \includegraphics[width=7.2cm]{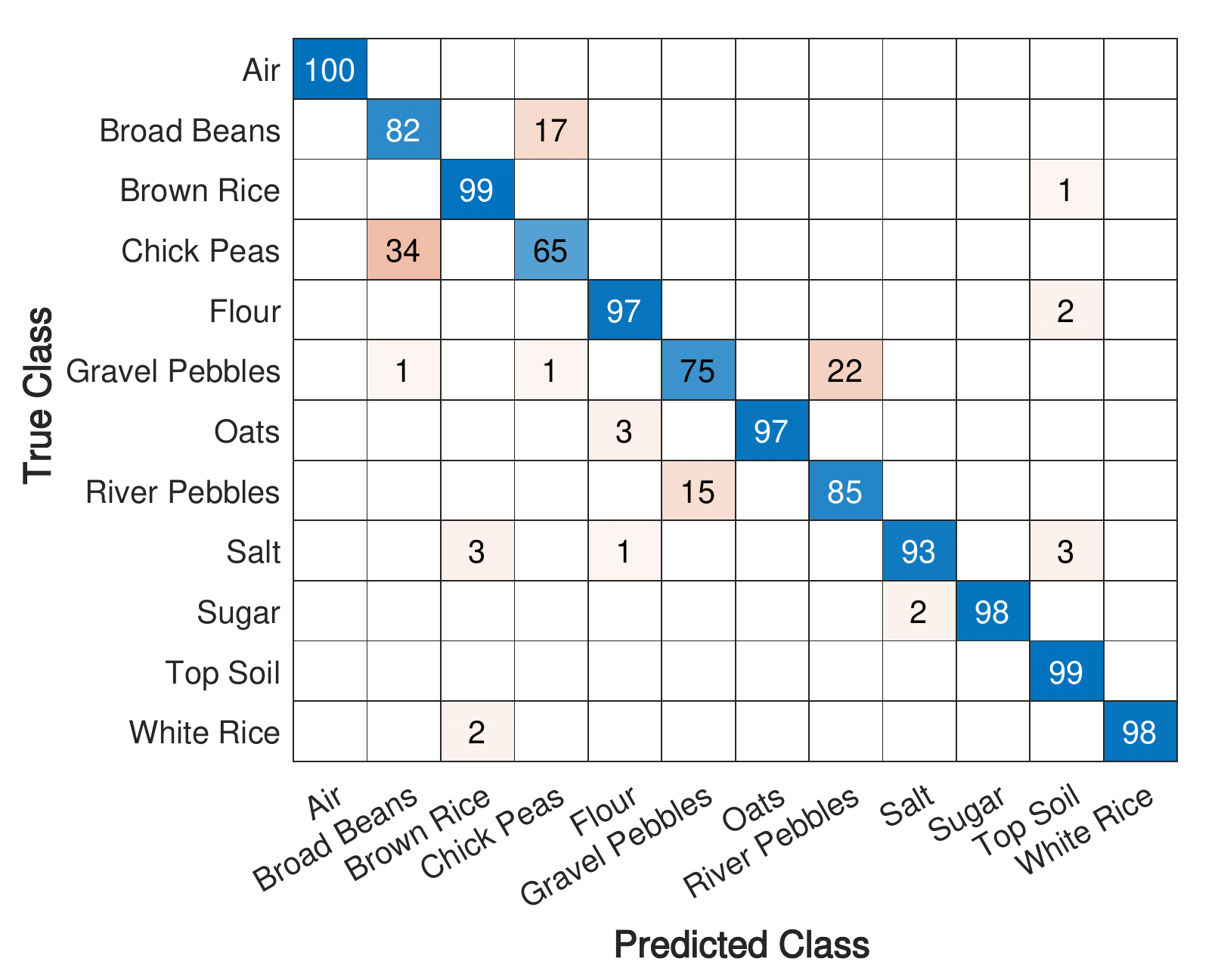}
        \caption{IMAGEVK-74}
    \end{subfigure}
    \caption{Classification confusion matrices for the two radar units (rounded to the nearest integer).}
\label{Figure06}
\end{figure*}

The classification confusion matrices for both of the IMAGEVK-74 and Walabot radars are shown in Fig. \ref{Figure06}. These results were generated using a 10-fold cross validation testing scheme that ensured that all classes were sampled equivalently. On the other hand, the online classification results in the next section employed completely unseen samples that were collected during the online experiments. 

These offline testing results clearly show the power of both radar units in predicting the different materials based on the collected offline radar signals. In general, the IMAGEVK-74 unit had an average of $90.66\pm11.40\%$ for the main diagonal of the confusion matrix while the Walabot had an average of $92.33\pm5.49\%$ for its main diagonal of confusion matrix. While these results demonstrate the lower-frequency Walabot was more accurate, the off-diagonal values within the two confusion matrices indicate that the higher-frequency IMAGEVK-74 made more `reasonable' confusions between the material classes than the Walabot. In this regard, one can clearly see that the IMAGEVK-74 confused broad beans with chick peas (both legumes with broadly similar surface texture) and gravel pebbles with river pebbles (both rocks). On the other hand, the Walabot made much smaller size confusions but these were spread more evenly across the different classes. 

\begin{figure}
    \begin{subfigure}{0.27\textwidth}
        \centering
        \includegraphics[width=8cm]{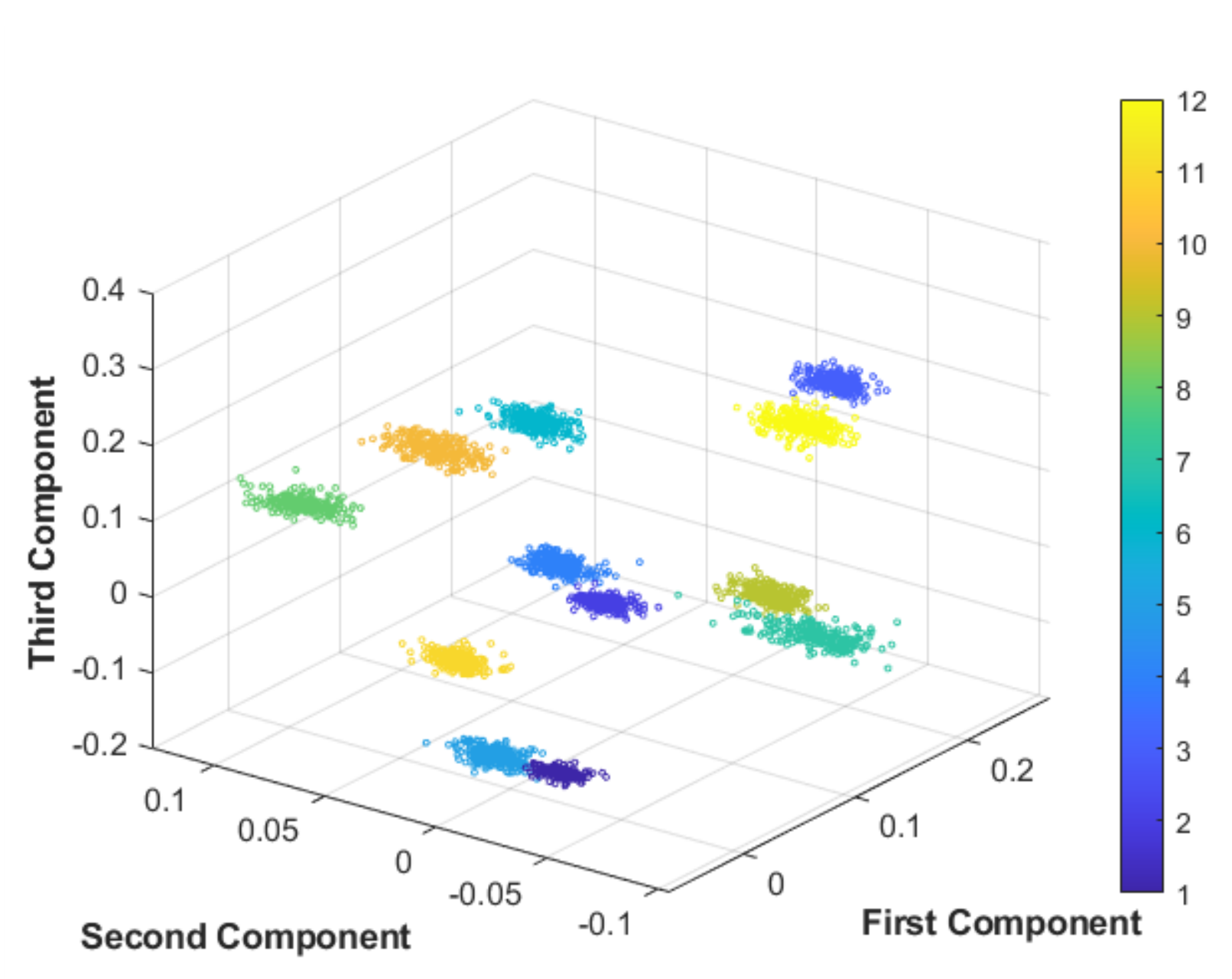}
        \caption{Walabot}
    \end{subfigure}
    \begin{subfigure}{0.27\textwidth}
        \centering
        \includegraphics[width=8cm]{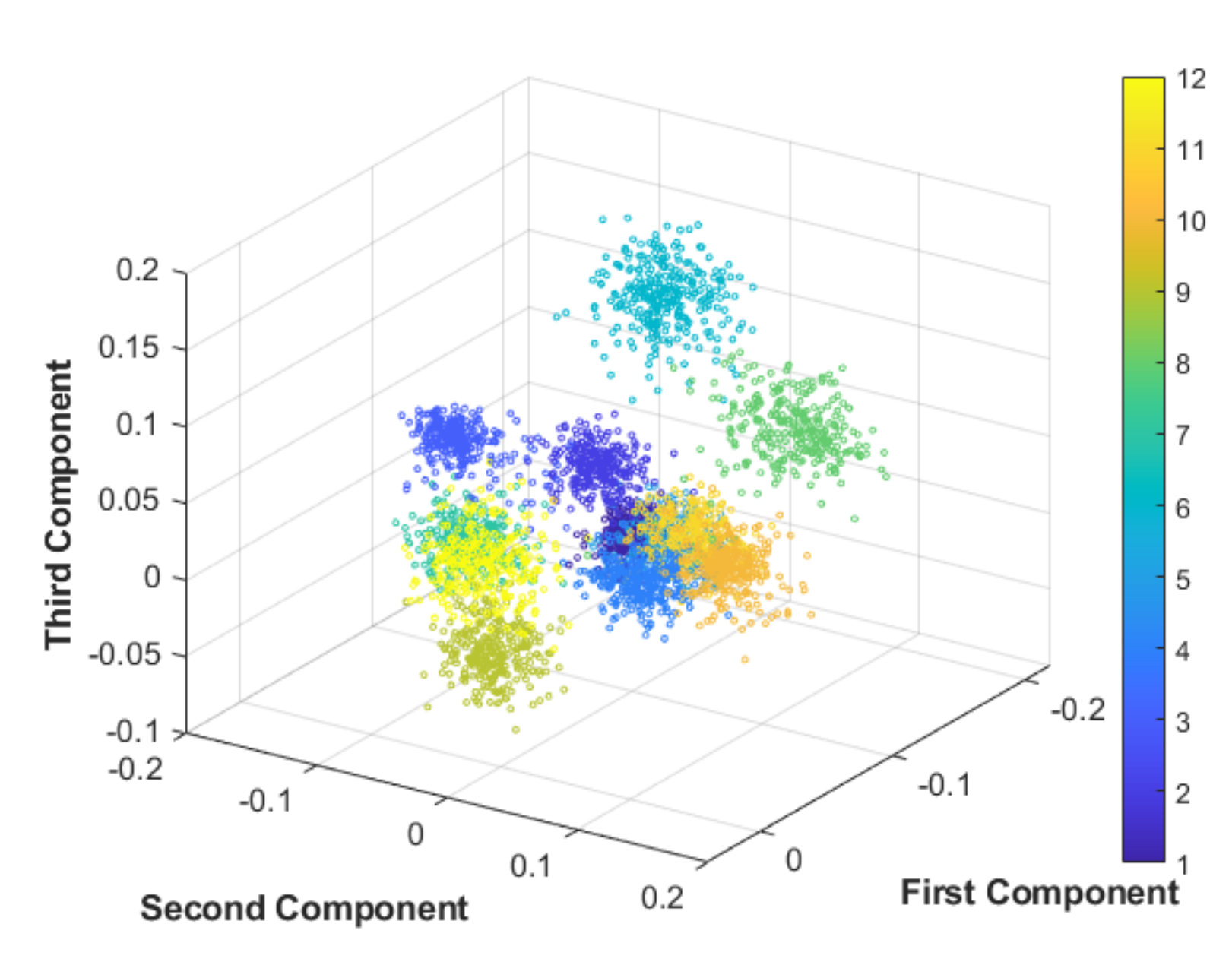}
        \caption{IMAGEVK-74}
    \end{subfigure}
    \caption{Scatter plots for the first 3 components out of the 11 generated components after SR projection for the two radar units. By considering more features from the IMAGEVK-74 scatter, the classification pipeline was able to further clarify any confusion between the classes.}
\label{Figure07}
\end{figure}

The scatter plots of the first three components after SR projection were also analyzed, shown in Fig. \ref{Figure07}. These plots show that the Walabot features had much better separation than the IMAGEVK-74 for the purpose of material identification. This in turn encouraged further online analysis to study the suitability of the two radar units for real-time decision making. This is particularly important for implementing this in an automated system, such as detecting grasped objects, part selection or QA/QC in manufacturing, or analyzing soil/rock properties in agriculture or interplanetary robotics.


\subsection{Online Classification Results}

\begin{figure}
    \begin{subfigure}{0.5\textwidth}
        \centering
        \includegraphics[width=7cm]{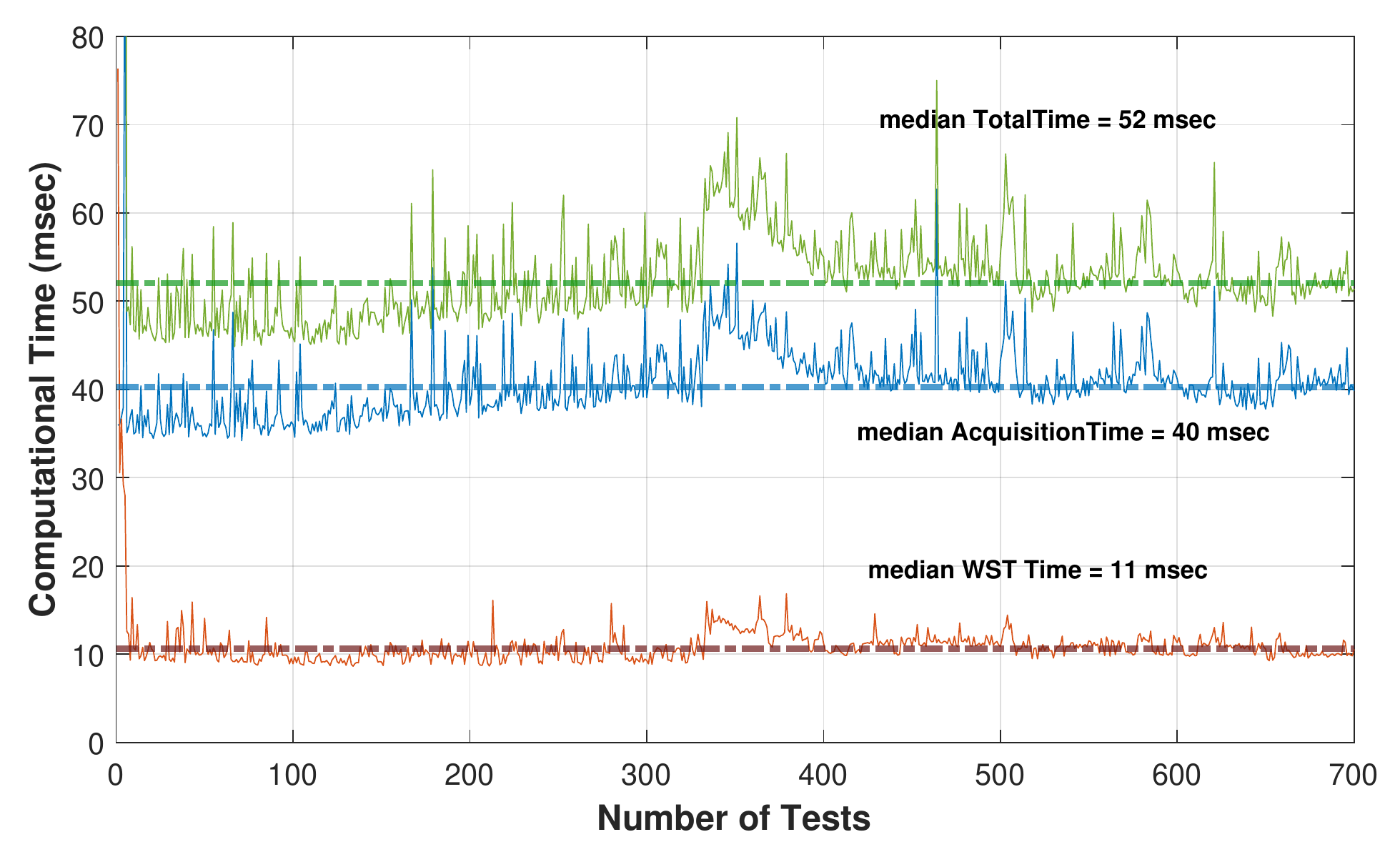}
        \caption{Walabot}
    \end{subfigure}%
    
    \begin{subfigure}{0.5\textwidth}
        \centering
        \includegraphics[width=7cm]{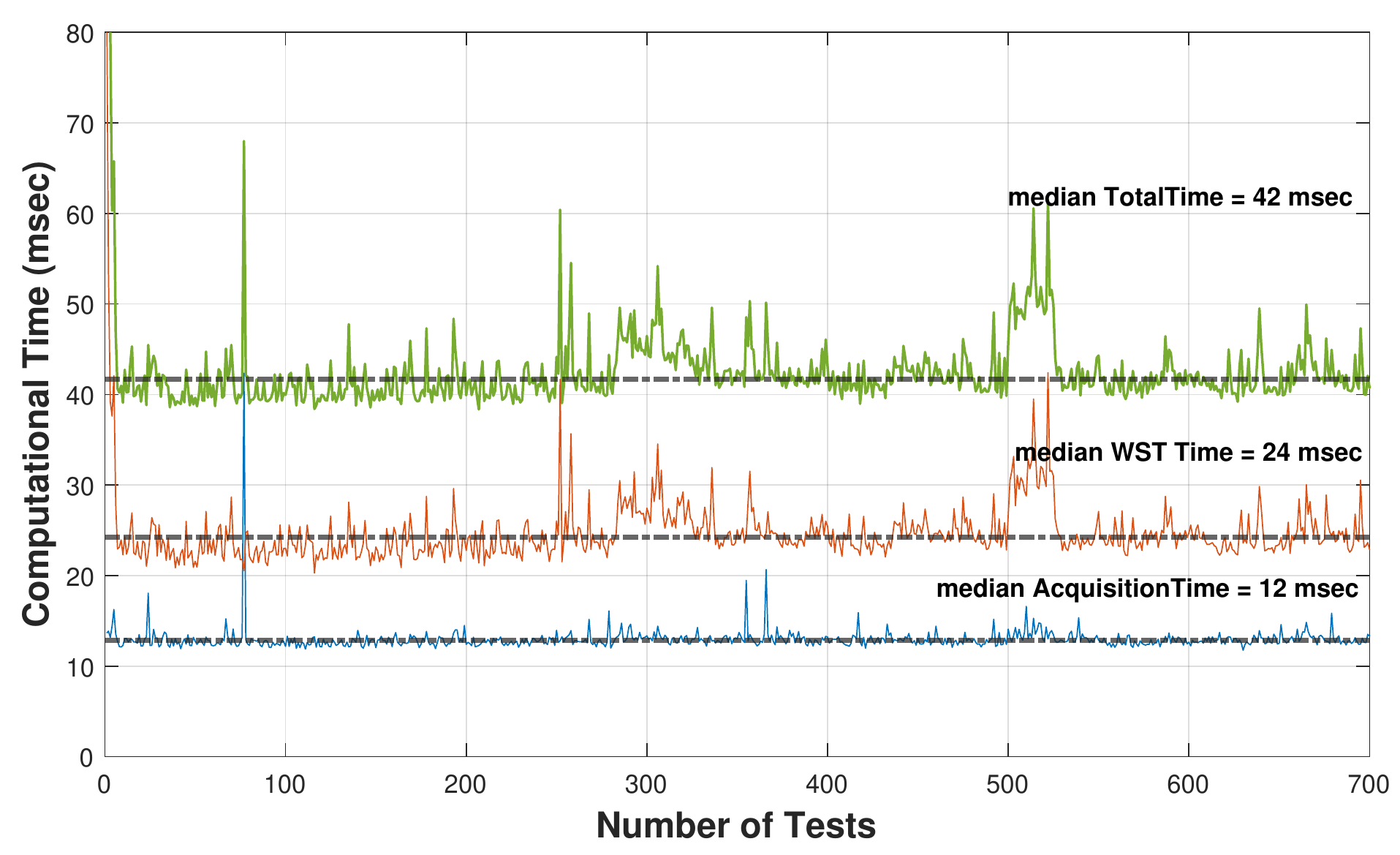}
        \caption{IMAGEVK-74}
    \end{subfigure}
    \caption{Computational time in ms for the processing pipelines.}
\label{Figure08}
\end{figure}

The online classification performances for the Walabot and IMAGEVK-74 processing pipelines were analyzed in separate experiments, measuring the real-time performance stability and computational costs. For both radar units, the testing samples were acquired during the online tests, i.e., completely unseen samples during the training phase. In the Walabot experiments, it can be seen from Fig. \ref{Figure08} that the majority of the computation time was spent acquiring the corresponding radar signals, which had a median time of $\approx 40$ ms. The median time for application of the WST was $\approx 11$ ms, and the remaining steps of feature normalization, SR projection and classification were all achieved within a median of just $\approx 1$ ms. Overall, the Walabot pipeline produced decisions on the corresponding materials within a median time of $\approx 52$ ms per iteration. On the other hand, the IMAGEVK-74 unit produced more signals (up to 400 raw time domain signals) but incurred a much lower acquisition time of $\approx 12$ ms. A further $\approx 24$ ms was required to process these signals with WST, and the overall median time per iteration was $\approx 42$ ms. The extra time consumed by the WST processing was due to the increased number of time-domain signals compared to the Walabot (400 vs 40 raw time-domain signals). These overall processing times of 40--52 ms are well-suited to real-time usage in many applications, allowing sampling rates around 20--25 Hz.

A video demonstration is also attached with this paper to demonstrate the performance of the two radar unit during online testing. This has been also made available here \url{https://www.youtube.com/watch?v=Mfohzvf7iuA}.

\section{Discussion}
Two radar units were considered for the problem of material identification with the wavelet scattering transform. The main aim of this study was to investigate and compare the suitability of the centimeter- vs. millimeter-wave radar units for this application, to understand if one of these units or frequency ranges is better suited for this task. While the literature has already demonstrated the effectiveness of similar radar units for this task individually, no previous comparisons of the two bands were found.

Our results suggest that the 6.3--8 GHz Walabot radar produced more accurate results during offline classification ($92.33\%$ vs $90.66\%$), with lower confusion between materials. This was also demonstrated in the attached video supplementary file with this paper. It is hypothesised that the lower wavelengths of this unit make it less sensitive to surface texture at this scale, resulting in a greater signal that relates to the material properties themselves (somewhat akin to the use of much lower frequencies for ground penetrating radar in geological sampling).

On the other hand, while classification results for the IMAGEVK-74 pipeline were slightly worse, the confusions made by this pipeline were perceived as less `random' by the authors. Examples include confusion between chickpeas and broad beans (both legumes), or gravel pebbles with river pebbles (both rocks). Depending on the application, this may be better or worse, especially combined with other sensing modalities. Overall, both units were utilized to accurately discriminate between the different materials, with the Walabot showing better robustness.

In terms of the computational costs for these two classification pipelines, our analysis showed that the Walabot pipeline took a median of 52 ms/decision while the IMAGEVK-74 required 42 ms. While both numbers appear suitable for online material identification applications, the IMAGEVK-74 was much faster in acquiring the signals, despite collecting far more samples than the Walabot. The longer processing time in the WST stage for the IMAGEVK-74, due to the larger input size, could perhaps be offset with optimisation of the code and/or computational improvements. Regardless, the units allow sampling in the order of 20--25 Hz, which is suitable for applications like humanoid grasping and remote material sensing applications or interplanetary robotics.

\section{Conclusions}

 
The aim of this paper was to investigate the suitability of centimeter- vs. millimeter-wave radar units for material identification, by using both units with the same pattern recognition pipeline. Given the Deep WST's ability to provide low-variance representations of the underlying signals, WST was hence suggested for feature extraction in this pipeline. Our research found that both units can be used for material identification, although the Walabot pipeline performed better in the online testing and had less confusion across classes in the offline tests. Overall, the IMAGEVK-74 unit had an average of $90.66\pm11.40\%$ for its main diagonal of the confusion matrix, while the Walabot had an average of $92.33\pm5.49\%$. While the Walabot captures fewer signals than the IMAGEVK-74 (due to fewer antenna pairings), it can be used to reliably solve this problem, at least in the context of the present experiments and materials evaluated in this study. Given the high accuracy for single-material classification, further work is suggested in mixed material samples, as will occur in applications and interplanetary scenarios, where identification of the mixed components may be of interest.




\section*{ACKNOWLEDGMENT}
The authors would like to acknowledge the support of the Avik Santra for providing feedback on \cite{Avik01} and \cite{Avik00}. Thanks also extends to Stephen Leone from mini-circuits for their support with the IMAGEVK-74 API and purchasing.


%

\ifCLASSOPTIONcaptionsoff
  \newpage
\fi

\end{document}